\documentclass[useAMS,usenatbib]{mn2e}
\usepackage[a4paper,colorlinks=True,citecolor=blue]{hyperref}
\usepackage{amssymb,esint}
\usepackage{graphicx}
\usepackage{epstopdf}
\usepackage{mathptmx}
\newcommand{\bea}{\begin{eqnarray}}
\newcommand{\eea}{\end{eqnarray}}
\newcommand{\beq}{\begin{equation}}
\newcommand{\eeq}{\end{equation}}

\newcommand{\Lam}{\ensuremath{\Lambda}}
\newcommand{\sm}{\small}

\title[The nature of voids II]{The nature of voids: II. Tracing underdensities with biased galaxies}
\author[S. Nadathur and S. Hotchkiss]{S. Nadathur$^{1}$\thanks{seshadri.nadathur@helsinki.fi} \& S. Hotchkiss$^{2}$\\
$^1$Department of Physics, University of Helsinki and Helsinki Institute of Physics, P.O. Box 64, FIN-00014, University of Helsinki, Finland\\
$^2$Department of Physics and Astronomy, University of Sussex, Falmer, Brighton, BN1 9QH, UK\\
}

\begin{document}

\date{\today}

\pagerange{\pageref{firstpage}--\pageref{lastpage}}

\label{firstpage}

\maketitle

\begin{abstract}
We study how the properties of cosmic voids depend on those of the tracer galaxy populations in which they are identified. We use a suite of halo occupation distribution (HOD) mocks in a simulation, identify voids in these populations using the {\sm ZOBOV} void finder and measure their abundances, sizes, tracer densities, and dark matter content. To separate the effects of bias from those of sampling density, we do the same for voids traced by randomly down-sampled subsets of the simulation dark matter particles. At the same sampling density, galaxy bias reduces the total number of voids by $\sim50\%$ and can dramatically change their size distribution. The matter content of voids in biased and unbiased tracers also differs. Deducing void properties from simulation therefore requires the use of realistic galaxy mocks. We discuss how the void observables can be related to their matter content. In particular we consider the compensation of the total mass deficit in voids and find that the distinction between over- and under-compensated voids is not a function of void size alone, as has previously been suggested. However, we find a simple linear relationship between the average density of tracers in the void and the total mass compensation on much larger scales. The existence of this linear relationship holds independent of the bias and sampling density of the tracers. This provides a universal tool to classify void environments and will be important for the use of voids in observational cosmology. 
\end{abstract}

\maketitle



\section{Introduction}
\label{section:intro}

Several recent studies have advocated the use of large underdensities in the matter distribution, known as cosmic voids, which can be important tools for cosmology in a number of different contexts. These include using voids to test modified gravity scenarios \citep{Li:2012,Clampitt:2013,Cai:2014,Zivick:2014}, to probe dark energy \citep{Ryden:1995,Lee:2009,Bos:2012,Sutter:2015a,Pisani:2015}, or to detect the effects of warm dark matter \citep{Tikhonov:2009,Yang:2014}. These tests make use of observables such as the abundances, size distributions, shapes and radial density profiles of voids. This information will be available for very large numbers of voids in future large-scale survey data such as Euclid \citep{Euclid:DSR}.

Weak lensing by voids \citep{Krause:2013,Melchior:2014}, and their possible gravitational effects on the CMB via the integrated Sachs-Wolfe (ISW) effect have also been studied \citep{Granett:2008ju,Cai:2013ik,Hotchkiss:2015a,Planck:2015ISW}. In principle for rare voids such studies probe the extreme tails of the distribution of fluctuations of the gravitational potential so may provide unique insights. Indeed the case of possible ISW imprints of voids is particularly intriguing, since theoretical considerations and simulation results show that any such detection would be in strong tension with the standard \Lam CDM cosmology \citep{Nadathur:2012,Flender:2013,Cai:2013ik,Hotchkiss:2015a,Aiola:2015}.

Several catalogues of voids found in galaxy redshift surveys have already been published \citep{Pan:2012,Sutter:2012wh,Nadathur:2014a}, and more will be available from future data releases. These catalogues can be used for the various studies described above. However, since the dark matter distribution is not directly observable, the comparison of observations with theoretical expectation necessary for precision cosmology requires a detailed understanding of how biased galaxies actually trace matter underdensities.

A common theoretical model of voids is that based on the excursion set formalism \citep{Sheth:2003py}, which can be modified to account for the bias of galaxy tracers \citep[e.g.][]{Furlanetto:2006}. However, this model assumes that voids can be described as spherical underdensities surrounded by higher-density walls that have reached shell-crossing. Recent results \citep{Falck:2014,Nadathur:2015b} show that this assumption does not hold for the most commonly used watershed void-finders, and more generally may not hold for any practical void-finding technique. Instead voids in simulation and observation correspond to a much more general set of density minima of varying depths.

This means that the development of cosmological predictions for void observables must be guided by and calibrated on simulation results. To properly account for the effects of bias in the galaxy distribution, these simulations should be populated with realistic mock galaxy populations, for example by using a halo occupation distribution \citep[HOD; e.g.][]{Berlind:2002} model. However, a number of recent studies of voids \citep[e.g.][]{Chan:2014,Zivick:2014,Sutter:2015a,Pisani:2015} do not follow this approach. Instead these authors randomly down-sample the dark matter particle output of simulations to the same mean number density as that of galaxies in various surveys, and use these particles as tracers of voids. This takes into account effects relating to the sampling density of tracers. However, the clustering properties of sub-sampled dark matter and galaxies are in general expected to differ, which should also affect voids. The justification for such a procedure has therefore relied on the results of \citet{Sutter:2014b}, who suggest that tracer bias has a smaller effect on void properties than sampling density and can therefore be ignored.

This result is however hard to reconcile with the expected dependence of the halo mass function on the background matter density on larger scales \citep{Sheth:2002}. \citet{Gottloeber:2003} find that the numbers of high- and intermediate-mass halos in voids is sharply suppressed, and \citet*{Alonso:2015} also find a strong dependence of the halo mass function on the density environment. This means that properties such as the sizes, abundances and density profiles of voids traced by galaxies in such halos should be expected to be very different to those traced by unbiased dark matter. The picture is further complicated by the fact that the halo abundance within voids is also expected to be cosmology-dependent \citep{Lee:2012,Neyrinck:2014}. 

We therefore revisit this question in this paper. We make use of a suite of different HOD models to generate biased galaxy tracer populations in an $N$-body simulation, and also use sets of sub-sampled dark matter particles with the same number densities as controls. We use these sets of tracers to identify voids using the {\sm ZOBOV} watershed void-finding algorithm, and compare their properties. We show that whereas the mean sampling density determines the minimum resolvable void size, at the same sampling density differences in the tracer bias produce large differences in the void size distribution, especially at large void sizes, as well as affecting the total number of detected voids and the densities within voids. This underscores the importance of using appropriate HOD models in simulations when obtaining predictions for void observables.

We then investigate the distribution of dark matter within galaxy voids. The minimum tracer density within each void, $n_\rmn{min}$, varies significantly over the population and we find that it can be used to predict the minimum dark matter density within the void. The nature of this relationship however depends strongly both on the tracer bias and the sampling density. The stacked average density profile of voids is also of interest, as it could potentially be a sensitive test of cosmology (\citealt*{Cai:2014}, \citealt{Barreira:2015}). Some studies (\citealt*{Hamaus:2014a}, \citealt{Nadathur:2015a}) have also suggested the possibility of a universal density profile, focussing on the variation of the average profile with the mean void size. We show that the average profile also depends strongly on $n_\rmn{min}$. 

An important related characteristic of voids is whether the total mass deficit within them is compensated or not. This is a key element in determining the environment within voids \citep{Hamaus:2014a}. Voids in which the central density deficit is not compensated out to large distances will correspond to gravitational potential hills, while over-compensated voids are smaller local density minima in larger-scale overdense regions, and will therefore correspond to regions of negative gravitational potential. Distinguishing between these two cases is therefore crucial for understanding void environments, and in particular for the use of voids in the cosmological tests described above. 

It has previously been suggested \citep{Ceccarelli:2013,Cai:2013ik,Hamaus:2014a} that such a distinction can be made on the basis of the void size. Instead we find that the transition from under- to over-compensation is dependent on \emph{both} the void size and minimum tracer density within the void, and is not well predicted by either property alone. However, we are able to identify a very simple linear relationship between the enclosed mass density contrast $\Delta$ at scales much larger than the void and the integrated average tracer number density within the void, $n_\rmn{avg}$. The existence of this linear relationship is found to be remarkably independent of the bias or sampling density $\overline{n}$ of the tracer population, with $n_\rmn{avg}=\overline{n}$ marking the transition between $\Delta<0$ (under-compensated) and $\Delta>0$ (over-compensated) for all tracer types. In contrast, the void size does not provide such a universal diagnostic of void compensation.

In Section~\ref{section:methods} below we describe the properties of the simulation, the construction of HOD galaxy samples and the void-finding algorithm. In Section~\ref{section:properties} we investigate the properties of these voids and lay out our results. We discuss their significance and conclude in Section~\ref{section:conclusion}. Some details of the stacked density profiles around voids and comparisons with previous results are presented in Appendix \ref{appendixA}.


\section{Methods}
\label{section:methods}

\subsection{Simulation}
\label{subsec:simulation}

We make use of $N$-body simulation data from the MultiDark simulation project \citep{Prada:2012}, in particular the MDR1 data release.\footnote{Publicly available at \url{www.cosmosim.org}.} MDR1 simulates a  \Lam CDM cosmological model with parameters $\left(\Omega_m,\Omega_\Lam,\Omega_b,h,n_s,\sigma_8\right)=\left(0.27,0.73,0.0469,0.7,0.95,0.82\right)$ using $2048^3$ dark matter particles in a cubic box of side $1\;h^{-1}$Gpc. The mass of each dark matter particle in the simulation is $8.7\times10^9\;h^{-1}M_\odot$, and it has a force resolution of $7\;h^{-1}$kpc. The initial conditions for the density evolution are set at redshift $z=65$ using the Zeldovich approximation.

\begin{table*}
\begin{minipage}{100mm}
\caption{Properties of the mock HOD samples used in this work. Bias values are approximate and intended as a guide only.}
\begin{tabular}{@{}cccccccc}
\hline
Sample name & $\log M_\rmn{min}$ & $\sigma_{\log M}$
& $\log M_0$ & $\log M_1^\prime$ & $\alpha$ & $\overline{n}$ & $b$ \\
& & & & & & $ (h^3\rmn{Mpc}^{-3})$ &\\
\hline
Main1 & 12.14 & 0.17 & 11.62 & 13.43 & 1.15 & $3.18\times 10^{-3}$ & 1.3 \\
Main2 & 12.78 & 0.68 & 12.71 & 13.76 & 1.15 & $1.16\times 10^{-3}$ & 1.4 \\
LOWZ & 13.24 & 0.62 & 14.32 & 13.20 & 0.93 & $2.98\times 10^{-4}$ & 2.0 \\
\hline\\
\end{tabular}
\label{table:HODsamples}
\end{minipage}
\end{table*}

Halos are identified in the simulation using the Bound Density Maximum (BDM) algorithm first described by \citet{Klypin:1997}, and as updated by \citet{Riebe:2013}. This algorithm detects as halos spherical regions around density maxima exceeding the virialization density threshold in a top-hat model of the growth of density fluctuations and removes gravitationally unbound particles from the halo. The virial radius of the halo, $R_\rmn{vir}$, is defined as the radius within which the density contrast $\Delta_\rmn{vir}(z)$ defined with respect to the background density $\overline{\rho}(z)$ exceeds the collapse threshold given by the approximation of \cite{Bryan:1998} -- at the redshift $z=0$ snapshot we use, this corresponds to $\Delta_\rmn{vir}=360$. The minimum resolved halo mass in MDR1 is $1.7\times10^{11}\;h^{-1}M_\odot$.

The creation of HOD mocks and other tracer samples for void identification is explained below; however, to measure the true density at void locations we make use of the underlying dark matter density field from the full resolution simulation output at redshift 0. This density field is determined on a $1024^3$ grid using a cloud-in-cell interpolation, and is then smoothed using a Gaussian kernel of width equal to one grid cell. The resulting density field has a sub-Mpc resolution, which is sufficient given the void sizes in our catalogues.

\subsection{HOD mocks and tracer samples}
\label{subsec:HOD}

We populate the halo catalogue with mock galaxies using the HOD model of \citet{Zheng:2007}, which assigns galaxies to a dark matter halo according to a distribution based on the halo mass $M$. Central and satellite galaxies are treated separately. According to this model, the mean occupation function of central galaxies is parameterized as
\beq
\label{eq:Ncen}
\left<N_\rmn{cen}(M)\right>=\frac{1}{2}\left[1+\rmn{erf}\left(\frac{\log M-\log M_\rmn{min}}{\sigma_{\log M}}\right)\right]\,,
\eeq
and the number of central galaxies in each mass bin follows a nearest-integer distribution. On the other hand the number of satellite galaxies follows a Poisson distribution with
\beq
\label{eq:Nsat}
\left<N_\rmn{sat}(M)\right>=\left<N_\rmn{cen}(M)\right>\left(\frac{M-M_0}{M_1^\prime}\right)^\alpha\,.
\eeq
Central galaxies are placed at the centre of their respective halos, while satellite galaxies are distributed through the halo with radial distances from the centre drawn from a random distribution based on a fiducial NFW mass profile \cite{NFW:1996,NFW:1997}.

The HOD distribution is thus characterized by 5 parameters: $M_\rmn{min}$, $M_0$, $M_1^\prime$, $\sigma_{\log M}$ and $\alpha$. We choose these parameter to match those determined in the literature from fits to the number density and clustering properties of observed galaxy samples.

Two of our mock samples, designated Main1 and Main2, are chosen to match volume-limited luminosity threshold samples from the Sloan Digital Sky Survey (SDSS) DR7 Main galaxy sample, with $r$-band absolute magnitude $M_r^\rmn{max}$ of $-20.5$ and $-21$ respectively \citep{Zehavi:2011}, with median redshift $z\sim0.1$. As these authors use a definition of the halo overdensity $\Delta_c=200$ with respect to the critical density which differs from the MultiDark definition above, we rescale the halo masses and virial radii from the simulation results using the concentration values provided and assuming an NFW profile. The large-scale linear bias $b$, defined by $\delta_g=b\delta$ where $\delta_g$ is the galaxy density contrast and $\delta$ that of matter, is $1.29$ for Main1 and $1.40$ for Main2.

Our third sample is chosen to match the properties of the Low Redshift (LOWZ) sample of galaxies from the SDSS BOSS data releases 10 and 11, with $0.15\lesssim z\lesssim0.43$, and is based on the HOD parameters obtained by \citet{Manera:2015}. Although the LOWZ sample in reality has a redshift-dependent mean sample density, we choose the parameter values for $\overline{n}=2.98\times10^{-4}\;h^3$Mpc$^{-3}$ and uniformly subsample our mocks down to achieve this mean density. \citet{Manera:2015} use halos within a simulation generated using the PTHalos method within second-order Lagrangian perturbation theory (2LPT) rather than full $N$-body simulations; their minimum halo mass is also too large to be able to resolve all populated halos, so they assign approximately 7\% of galaxies to dark matter particle positions rather than halos. These differences mean that our mock sample is unlikely to precisely match the clustering properties of the observed LOWZ galaxies; nevertheless it provides a sufficiently realistic realization of a distribution of galaxies with bias $b\sim2$.

Table~\ref{table:HODsamples} summarizes the parameter values and resulting average number density $\overline{n}$ and large-scale bias $b$ of our mocks. 

To complement these mock samples and to serve as controls, we also use three sets of randomly subsampled DM particles from the simulation output as tracers with which to identify voids. These samples are referred to as DM Main1, DM Main2 and DM LOWZ and have the same average tracer densities as the Main1, Main2 and LOWZ HOD samples respectively. However, as the random subsampling does not change the fundamental clustering properties of the DM field, these three samples are all unbiased ($b=1$). Finally, we also use three Poisson point samples, consisting of uniformly distributed random points with the same mean density as those of the mock galaxy tracers.

\subsection{Void finding and merging}
\label{subsec:voidfinding}

We make use of the {\sm ZOBOV} watershed void finding algorithm \citep{Neyrinck:2008} to identify voids in each of the tracer samples described above. {\sm ZOBOV} uses a Voronoi tessellation field estimator (VTFE) technique to reconstruct the tracer density field from a discrete distribution of particles. It then identifies local minima in this field and the watershed basins around them. These basins form a non-overlapping set of density depressions or voids.

At this stage, {\sm ZOBOV} can also merge neighbouring voids together according to the watershed principle to form a final hierarchy of voids and sub-voids. Depending on the choice of input parameters to control void merging, the resulting final void hierarchy and distribution of void sizes can vary widely: at one extreme one obtains percolation, with the entire simulation box or Universe being included in the largest void; at the other extreme there is no merging at all and the original voids are retained.

Unfortunately, the choice of how to control void merging is largely subjective and different options have been applied in the literature. \citet{Neyrinck:2008} suggests using the ratio of the lowest (VTFE-reconstructed) tracer number density along the void wall to the minimum VTFE density at its centre, relative to the same ratio for spurious voids in Poisson noise, to determine  the ``most probable extent". An alternative is to stop the growth of a void through merging if the minimum VTFE density along the watershed ridge separating it from a shallower potential sub-void, $n_\rmn{link}$, is greater than some pre-defined threshold. The {\sm VIDE} void finding toolkit \citep{VIDE:2015}, which is based on {\sm ZOBOV}, sets this threshold at $0.2$ times the mean tracer density $\overline{n}$. However, this value of 0.2 lacks any theoretical justification even when the tracers are dark matter particles in a simulation, because the recovered voids do not in any case correspond to the model objects described by \citet{Sheth:2003py} from which the canonical value for the enclosed density contrast $\Delta=-0.8$ derives \citep{Nadathur:2015b}. \footnote{Even if such a correspondence existed between the theoretical model and practical void-finding algorithm, there is no clear relationship between $n_\rmn{link}$ and $\Delta$ --- note that a restriction on $n_\rmn{link}<0.2\overline{n}$ does \emph{not} restrict the average density of the void to be less than 0.2 --- so this choice of threshold value would still be unmotivated.}

A further difficulty with the {\sm VIDE} threshold for merging is that it applies to the minimum tracer \emph{number} density along void walls. Therefore, when applied to differently biased tracer samples, the same numerical value for the threshold corresponds to rather different dark matter densities, and thus to different physical criteria for merging. Only for the special case where the threshold value is set equal to the mean sample density, i.e. neighbouring voids are merged together if $n_\rmn{link}<\overline{n}$, does this strategy result in the same physical consequences independent of the tracer bias.

\begin{table}
\centering
\caption{Strategies for controlling merging in the void hierarchy}
\begin{tabular}{@{}ccc}
\hline
Label &  \multicolumn{2}{c}{Criteria for merging:} \\
 & Link density & Density ratio \\
\hline
 VIDE & $n_\rmn{link}<0.2\overline{n}$ & unconstrained \\
Minimal & $n_\rmn{link}<\overline{n}$ & $r<2$ \\
Isolated & no merging & no merging \\
\hline
\end{tabular}
\label{table:merging}
\end{table}

In this paper we will compare three different strategies to control the merging of voids in the formation of the final hierarchy. The first strategy is that applied in the {\sm VIDE} algorithm, which has been used in several recent works \citep{Hamaus:2014a,Chan:2014,Zivick:2014,Sutter:2015a,Pisani:2015,Pollina:2015} studying void properties. The second strategy is simply to use the original output of density minima from {\sm ZOBOV} without any merging at all, which provides a set of non-overlapping voids defined purely by the topology of the tracer density field. We shall refer to the voids thus obtained as `Isolated' voids. 

The third strategy we consider allows merging of voids if the minimum tracer density on the separating ridge is below the mean. However, as this may still allow genuinely distinct structures to be classified as a single void, we implement an additional condition on the density ratio described above, i.e. we prevent the merging of two voids if the minimum density on the watershed ridge is more than twice the minimum density of the shallower void. This choice is based on the Poisson noise characteristics discussed by \citet{Neyrinck:2008}: it is designed to count neighbouring voids as distinct if the ridge separating them could not be an artefact of the discrete sampling of the density field. We refer to this as the `Minimal' choice of void merging criteria; it has also previously been used by \citet{Nadathur:2014a,Hotchkiss:2015a,Nadathur:2015a}. The three strategies for void merging are summarized in Table~\ref{table:merging}.

Most qualitative conclusions regarding the properties of voids in the simulation are independent of the particulars of the merging strategy chosen; some exceptions to this statement are highlighted below. However, quantitative results will inevitably depend on this choice. Since the `Isolated' option of preventing merging altogether provides a void definition that depends only on the topology of the tracer density field without further arbitrary parameters, it may be the one most easily described by a future theoretical model. Therefore, unless otherwise stated, the results presented in this paper relate to this case.

\subsection{Defining void properties}
\label{subsec:voidprops}

Based on the operational choices described above, we obtain a catalogue of voids within each of our tracer samples. Each void contains a number of tracer particles, and the total void volume is the sum of the volumes of the Voronoi cells of each of its member particles in the tessellation. We define the void effective radius, $R_v=(3V/4\pi)^{1/3}$, to be the radius of a sphere with the same volume as the void, although note that individual voids may be highly aspherical. The void member particle associated with the minimum VTFE-reconstructed tracer density (i.e., the largest Voronoi cell) is labelled the core particle, and this density is denoted by $n_\rmn{min}$. We previously showed \citep{Nadathur:2015b} that for subsampled dark matter tracers $n_\rmn{min}$ is tightly correlated with (but not the same as) the minimum dark matter density within the void, and inversely correlated with $R_v$, so that larger voids contain deeper density minima.

The centre of each void is defined as the point of intersection of the Voronoi cells of the void core particle and its three lowest-density mutually adjacent neighbours in the tessellation, which is the point within the void maximally distant from any tracer particles. Compared to the alternative volume-weighted barycentre definition often used in the literature, this definition is much better at locating the true dark matter density minimum within voids \citep{Nadathur:2015b}. Unlike the barycentre definition, it is also independent of the details of void merging and is less affected by shot noise. 

Given the total number of member tracer particles of a void and its total volume, it is also simple to define an integrated average tracer density for the void, $n_\rmn{avg}$. This density is obviously always greater than $n_\rmn{min}$; it is also well known \citep[e.g.][]{Nadathur:2014a,Nadathur:2015a} that the operation of the watershed algorithm means that for some voids $n_\rmn{avg}$ can be much larger than $n_\rmn{min}$ and even $>\overline{n}$ due to the inclusion of higher-density filaments, walls and halos within the watershed basin constituting a void.

\begin{figure*}
\begin{center}
\includegraphics[width=165mm]{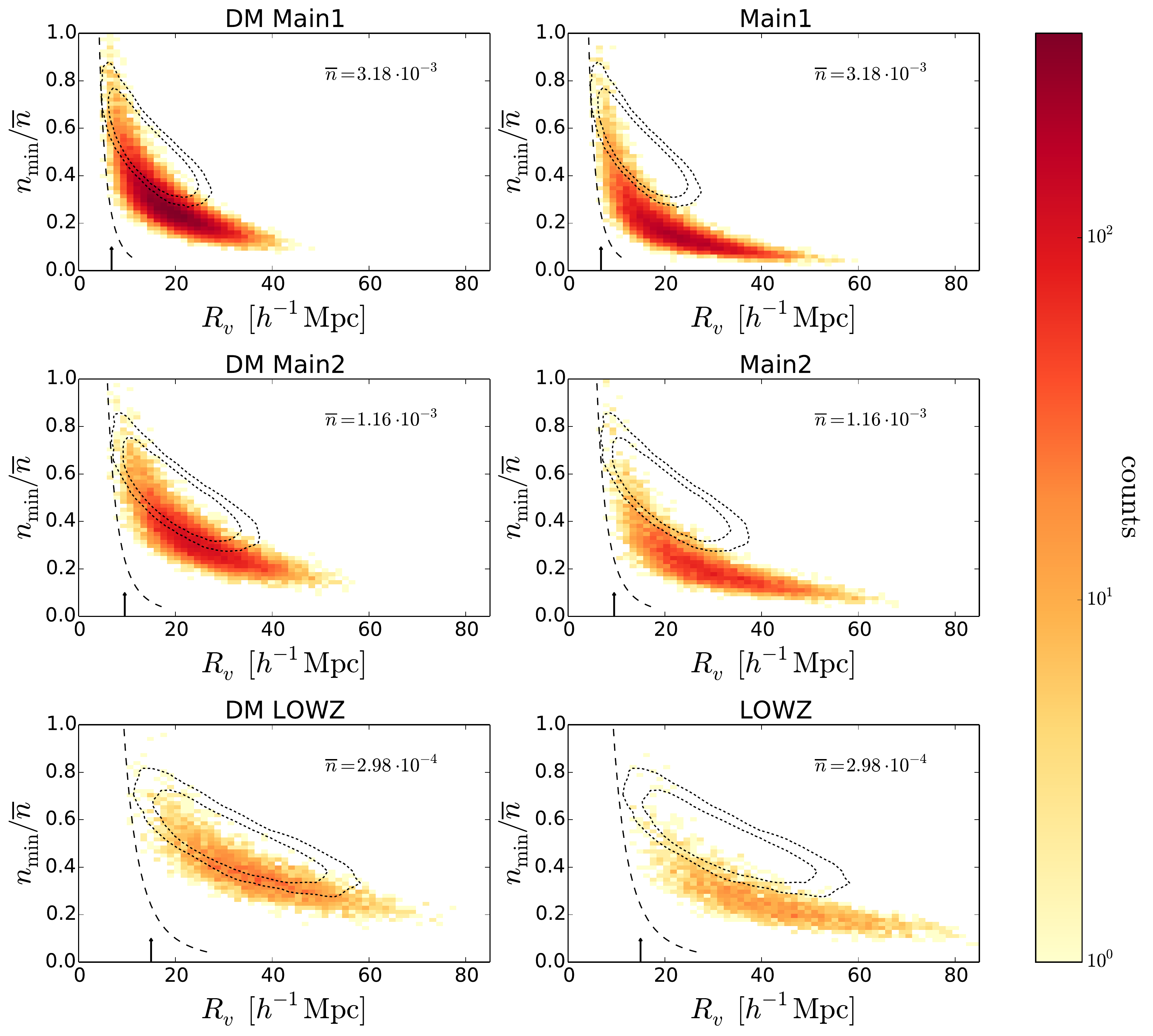}
\caption{Binned distribution of void minimum tracer densities $n_\rmn{min}$ and radii $R_v$, for all tracer populations as indicated. Plots in the left-hand column are for unbiased subsampled dark matter particle tracers, and those in the right-hand column are for the corresponding galaxy tracers with the same average number density $\overline{n}$. Dotted contours in each plot enclose $95\%$ and $99\%$ of all spurious voids found in uniform random distributions of points with the same $\overline{n}$ but no clustering. The dashed line indicates the minimum resolvable void size as a function of density, and the arrow shows the approximate mean interparticle separation $\overline{n}^{-1/3}$. Note that the number of bins remains the same in each plot, so bins cover larger ranges in the right column.} 
\label{fig:bananas}
\end{center}
\end{figure*}

The location of the centre together with $n_\rmn{min}$, $n_\rmn{avg}$ and $R_v$ constitute the core observable properties of a void, which can be defined from the tracer population alone.\footnote{To this list one could also add some measure of the asphericity and direction of alignment of the axes of aspherical voids, but we will not consider these properties in this work. The ratio of tracer densities at the void centre and walls and the full tracer density profile around the void are also in principle measurable, but can be complicated by survey boundary effects in observational data \citep{Nadathur:2014a,Nadathur:2015a}} For the same set of minima in the tracer density field, $n_\rmn{avg}$ and $R_v$ will differ depending on the details of the void merging criteria described above, but the location of the void centre and $n_\rmn{min}$ are independent of this choice. 

Using the gridded dark matter density field from the simulation output we can also measure the true density contrast $\delta=\rho/\overline{\rho}-1$ within the void and define the total enclosed density contrast within a sphere of radius $r$, centred at the void centre:
\beq
\label{eq:Delta}
\Delta(r)=\frac{3}{r^3}\int_0^r\left[\frac{\rho(y)}{\overline{\rho}}-1\right]y^2dy\,.
\eeq
Within the void, at $r<R_v$, values of $\Delta$ vary widely for voids of different radius and $n_\rmn{min}$  \citep{Nadathur:2015b}. In this paper, a property we consider is the compensation of the void underdensity on scales much larger than the void, for which we will use the value of $\Delta$ within a sphere three times the void radius, $\Delta(r=3R_v)$, as a diagnostic.

\begin{figure*}
\includegraphics[width=165mm]{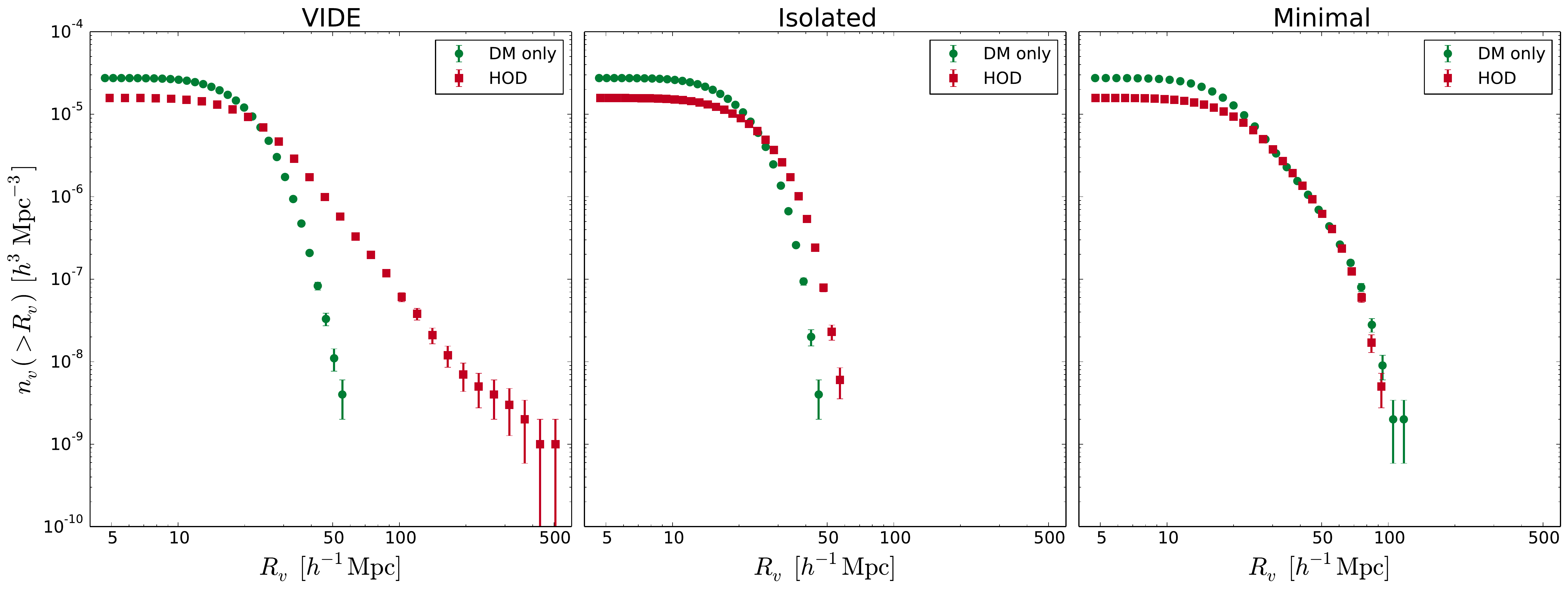}
\caption{The cumulative number density of voids with radius $>R_v$, for all voids in the unbiased DM Main1 sample, and for those in the Main1 HOD  galaxy mocks. The mean tracer density $\overline{n}$ is the same for these two samples, so differences in the void sizes are purely due to galaxy bias. The three panels correspond to three choices of the criteria to handle void merging described in Section~\ref{subsec:voidfinding} and Table~\ref{table:merging}. The bias of the HOD mocks decreases the total void abundance by $\sim50\%$ in all cases, and has a large effect on the shape of the distribution, especially for the merging criteria implemented in the VIDE toolkit. } 
\label{fig:nvR}
\end{figure*}


\section{Properties of voids}
\label{section:properties}

\subsection{Effect of tracer bias on void abundances and sizes}
\label{subsec:sizes}

Figure~\ref{fig:bananas} shows the distribution of void sizes $R_v$ and core tracer densities $n_\rmn{min}$ for voids each of the tracer samples described in Section~\ref{subsec:HOD}. The right hand plot in each row shows the distribution for one of the HOD mock samples (Main1, Main2 or LOWZ); on the left is the corresponding distribution for the voids found using the dark matter particles subsampled down to the same average tracer density $\overline{n}$. Therefore within each row, the sampling density remains the same but the bias increases. In the left-hand column, the bias remains the same (i.e. $b=1$) but the sampling density decreases. The dotted contours on each plot show the $95\%$ and $99\%$ C.L. contours for the equivalent distributions for voids found by the same algorithm in the random Poisson point samples: these contours therefore remain the same from left to right in a row. The dashed curve in each plot indicates the minimum resolvable void size at the given $\overline{n}$ \citep{Nadathur:2015b}. To clarify the discussion below, for the populations shown in this figure no void mergers have been allowed (corresponding to `Isolated' voids described in Section~\ref{subsec:voidfinding}). However, the qualitative features of the figure do not depend on this choice.

Several effects of tracer bias can be discerned from this figure. As discussed in \citet{Nadathur:2015b}, the characteristic `banana' shape of the distribution is naturally common to voids in all samples. However, at a quantitative level void sizes and core tracer densities depend strongly on the tracer bias as well as the mean tracer density. Although the minimum void size is determined by the resolution limit set by the sampling density $\overline{n}$, voids in mock galaxy samples span a larger range of sizes than those in corresponding sub-sampled dark matter populations. Even more importantly, galaxy voids are emptier of tracers than dark matter voids, as evidenced by the distribution of minimum tracer densities $n_\rmn{min}$ extending to lower values in the right column. 

\begin{figure*}
\includegraphics[width=165mm]{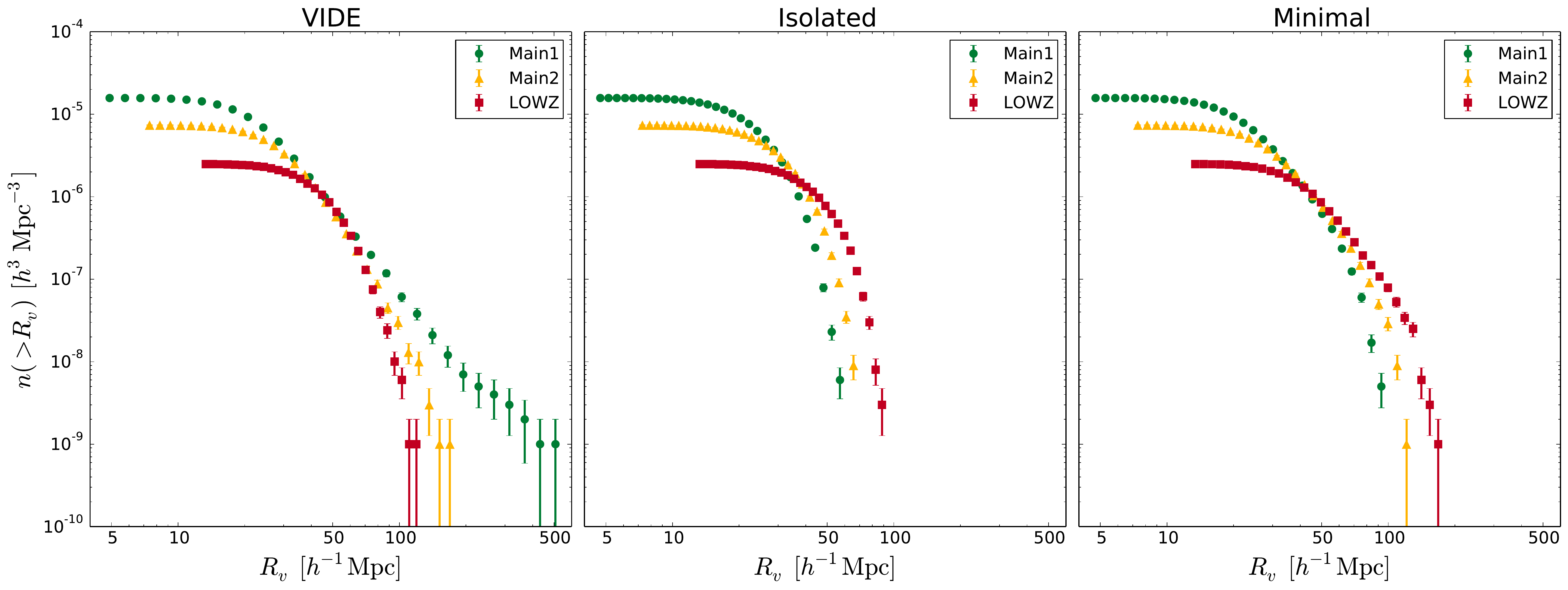}
\caption{The cumulative number density of voids with radius $>R_v$, for all voids in the three different HOD  galaxy mocks. The  panels correspond to the different choices of the criteria to handle void merging described in Section~\ref{subsec:voidfinding}. For the `Isolated' and `Minimal' merging criteria sparser and more highly biased galaxies consistently trace larger voids, as expected, but the situation is reversed for the VIDE choice of void merging. } 
\label{fig:nvR-HOD}
\end{figure*}

This difference in $n_\rmn{min}$ distributions has important consequences for void merging if the criterion for allowing merging is based --- as in the VIDE case discussed in Section~\ref{subsec:voidfinding} --- on an arbitrary number density threshold. To see this, note that according to the VIDE choice of merging criteria, a necessary but not sufficient condition for two neighbouring voids to be merged together as part of a larger void is that both must have $n_\rmn{min}<0.2\overline{n}$. As can be seen by comparing columns of Figure~\ref{fig:bananas}, a much larger fraction of galaxy voids satisfy this condition than dark matter voids. This means that, if merging is controlled by such a simple threshold alone, void mergers will be much more common for voids in galaxy distributions than in sub-sampled dark matter. Similarly, mergers would be more common between Main1 voids than those in LOWZ. This will necessarily increase the discrepancy between the observable properties of voids in the two populations, as we show later.

It is also noticeable that at each sampling density $\overline{n}$ the distribution of voids in sub-sampled dark matter populations shows a greater overlap with that of spurious Poisson voids than that of the corresponding voids in biased galaxy tracers. An intuitive explanation for this is that random sub-sampling increases shot noise, obscuring the true clustering properties of dark matter and increasing spurious void detections. For galaxy samples, the decrease in $\overline{n}$ is accompanied by increasing bias and therefore enhanced clustering, so they should be less affected by shot noise. However, we caution that just because a void identified in the simulation has $\left(R_v,n_\rmn{min}\right)$ values compatible with those found for spurious Poisson voids it does not necessarily follow that this void is itself an artefact of noise! While results for individual voids vary, we find that statistically speaking voids within the overlap region still tend to correspond to true matter underdensities. For this reason we do not apply any of the conservative cuts to the void population previously used in the literature (see \citealt{Nadathur:2015b} for a discussion).

Finally, we also find that the overall abundance of voids decreases significantly when the tracer bias is increased at fixed sampling density $\overline{n}$. In fact for each of the sampling densities considered, the total number of voids in unbiased sub-sampled dark matter particle populations exceeds that in biased galaxy populations by $\sim50\%$. This is significantly larger than the variation in void abundance due to changes in cosmological model described by, e.g., \citet{Zivick:2014,Sutter:2015a,Pisani:2015}. We conclude that predictions for this key observable that are obtained from calibration with simulations that do not include realistic mock galaxy populations are not observationally relevant.

To demonstrate this difference in void abundances more clearly, in Figure~\ref{fig:nvR} we plot the cumulative number function of voids as a function of their size for the DM Main1 and Main1 HOD samples, which both have the same sampling density $\overline{n}=3.18\times10^{-3}\,h^3\rmn{Mpc}^{-3}$. Since the three choices of void merging strategy produce different results, they are shown separately, as indicated by the titles of each panel. The overall deficit of galaxy voids is clearly visible and is due to a lack of small-to-intermediate sized voids, in the range $5\lesssim R_v\lesssim20\,h^{-1}$Mpc. 

At large $R_v$, the number density of galaxy voids is strongly dependent on the void merging criteria. The biggest difference between the DM-only and HOD distributions at large $R_v$ is obtained for the VIDE choice of merging threshold, for the reasons explained above. Note that the VIDE toolkit has been used by a large number of recent void studies which describe the properties of voids using down-sampled dark matter tracers as a substitute for galaxies, e.g. \citet{Sutter:2014b,Chan:2014,Zivick:2014,Sutter:2015a,Pisani:2015}. In fact, for the Main1 HOD sample the VIDE criterion leads to percolation through the simulation box, since the largest void has $R_v>500\,h^{-1}$Mpc. Such percolation only appears when void merging is controlled using a simple density threshold alone, as is the case for VIDE. It is strongly dependent on the choice of the arbitrary density threshold, as well as the sampling density and bias of the tracers. Such percolation may generally be considered an undesirable feature in choosing a void definition. 

Note that since the topology of the tracer density field is bias-dependent, even for the case of `Isolated' voids with no merging the void size distribution differs between HOD mocks and dark matter tracers at the same sampling density. For the `Minimal' merging criteria $n_v(>R_v)$ is independent of tracer bias for large $R_v$ but significant differences remain in the distribution at small and intermediate scales, and in the total void abundance.

Figure~\ref{fig:nvR-HOD} shows the differences in the void size distribution over the three mock galaxy samples. Intuitively, increasing tracer bias should shift the distribution to larger void sizes as well as decreasing the total number of voids: more highly biased galaxies live in higher mass halos and are therefore confined to higher-density environments, resulting in more large voids, while the lower sampling density $\overline{n}$ inhibits the ability to resolve small voids. Indeed this is what is seen in the middle and right panels of Figure~\ref{fig:nvR-HOD}, for the `Isolated' and `Minimal' merging choices. However, this intuitive picture is inverted for the VIDE case, for which the maximum void size in each sample instead decreases with increasing galaxy bias. This is because of the fixed threshold on tracer number density for merging, which has different physical consequences in each galaxy sample. It provides another reason to disfavour this choice of merging.

\begin{figure*}
\includegraphics[width=170mm]{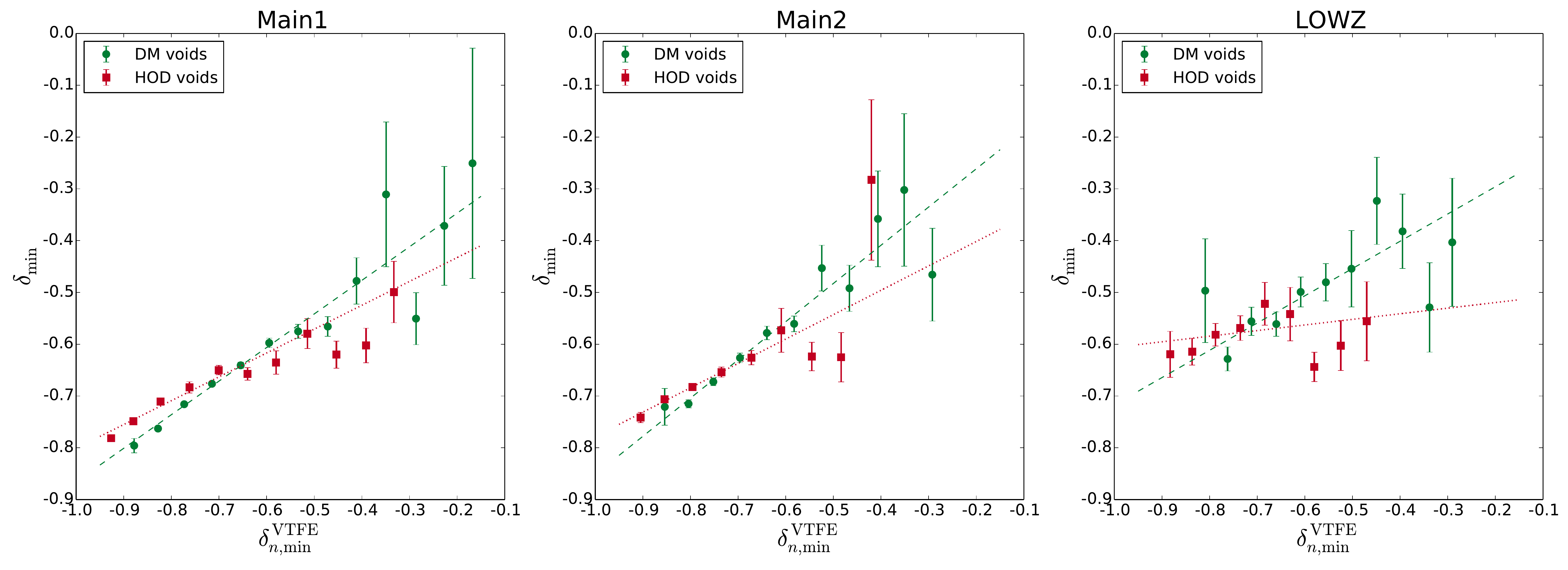}
\caption{The binned average dark matter density at the void centre as a function of the minimum tracer number density determined from the VTFE reconstruction, for voids in different tracer populations. In each plot, green circles are for voids traced by sub-sampled dark matter particles and red squares are for voids in the corresponding biased galaxy mocks. Error bars indicate the $1\sigma$ errors in the mean. The best straight-line fits to the unbinned data are shown with dashed (dotted) lines for dark matter (HOD) voids in each case. At low densities, voids in HOD galaxy mocks show smaller values of $\delta_\rmn{min}^\rmn{VTFE}$ due to biasing effects. Shallow voids in the galaxy distribution are less affected by shot noise due to subsampling and so trace deeper matter underdensities.} 
\label{fig:biases}
\end{figure*}

\subsection{Dark matter within voids}
\label{subsec:DMinvoids}

Although it is the dark matter content of voids that is of primary interest for cosmology, in practice it is only the tracer number densities at void locations that can be directly observed. These two quantities are obviously distinct for galaxy tracers, but \citet{Nadathur:2015b} showed that the tracer number density is not the same as the true matter density \emph{even when the tracers are themselves a random subset of all dark matter particles}. This is especially true when the number densities are measured by simply counting tracer particles in a volume around special locations within the void, such as the void centre. In fact, since as described in Section~\ref{subsec:voidprops} the void centre is by construction far from all tracers, naively reconstructing the tracer number density within a small enough volume around the centre returns a tracer number density contrast $\delta_n=-1$ for practically all voids. As the true matter densities within voids vary quite widely, this means that this naive measure of the tracer number density is not useful.

However, the VTFE-reconstructed tracer density is a better indicator of the true matter density \citep{Nadathur:2015b}. This is because it is less subject to sparse sampling effects (indeed this is the main reason why the VTFE density estimator is used in the first place). We find that the minimum or core VTFE-reconstructed tracer density contrast within the void, $\delta_{n,\rmn{min}}^\rmn{VTFE}\equiv n_\rmn{min}/\overline{n}-1$, is a good predictor of the underlying matter density contrast $\delta_\rmn{min}$ at the void centre.\footnote{Strictly speaking the correspondence between the void centre and the dark matter density minimum is statistical and may not hold for an individual void. However, any other location within the void that can be deduced on the basis of the tracer positions alone will on average always have a higher matter density.} Figure~\ref{fig:biases} shows measurements of $\delta_\rmn{min}$ for voids binned by their values of $\delta_\rmn{min}^\rmn{VTFE}\equiv n_\rmn{min}/\overline{n}-1$, for each of our tracer samples. There is a simple linear relationship between the two, as evidenced by the goodness of the straight-line fits to the data. However, note that $\delta_\rmn{min}^\rmn{VTFE}\neq\delta_\rmn{min}$ even when using dark matter tracers, and $\delta_\rmn{min}^\rmn{VTFE}\neq b\delta_\rmn{min}$ for voids in the galaxy mocks. 

Some qualitative aspects of Figure~\ref{fig:biases} can be understood in terms of the role of bias and subsampling. At low values of $\delta_\rmn{min}^\rmn{VTFE}$, voids in galaxy mocks on average correspond to shallower dark matter underdensities than do voids in the subsampled dark matter tracers with the same $\delta_\rmn{min}^\rmn{VTFE}$. This is a result of the overall bias of the galaxies and the suppression of higher mass halos in low density regions --- such regions therefore look even emptier in the galaxy number density. However, as $\delta_\rmn{min}^\rmn{VTFE}$ increases towards zero, the situation reverses: this is because subsampling of the dark matter particles increases the effect of shot noise and means that shallow depressions in the tracer density are less likely to correspond to true matter underdensities in this case than for the biased galaxy mocks. This is also consistent with the picture from Figure~\ref{fig:bananas}, where voids in subsampled dark matter are much more likely to overlap with those in equivalent Poisson point sets.

These qualitative features are common to voids in each of the Main1, Main2 and LOWZ samples. However, at a more quantitative level a cursory search did not reveal a simple way to relate the slope and intercept of the straight-line fits to the tracer bias and mean number density. This suggests that in order to assess the depth of the matter underdensity in voids in observational data, calibration using realistic HOD mocks is required on a case-by-case basis.

\begin{figure*}
\includegraphics[width=180mm]{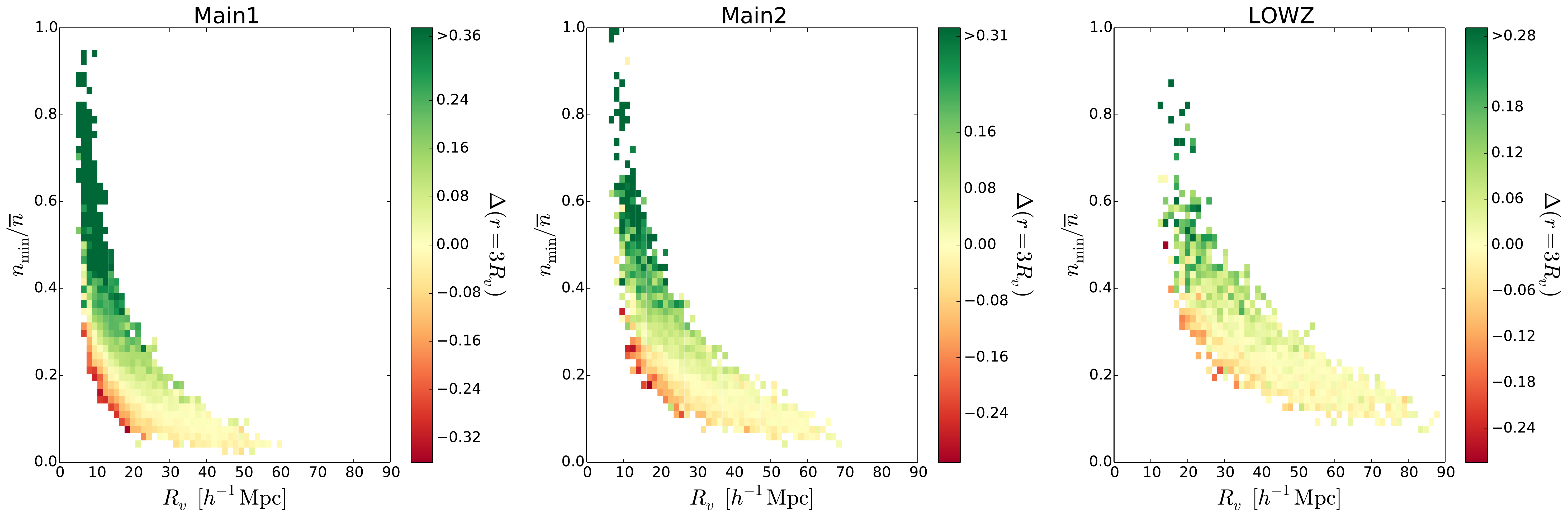}
\caption{The variation of void compensation as a function of minimum tracer density and void size. The colourbar indicates the binned average values of the total matter density contrast enclosed within a sphere of radius $3R_v$ centred at the void locations. The same trends are seen for voids in all three galaxy mocks: small and shallow voids are on average over-compensated, while deep and large voids are under-compensated, but the variation depends on a combination of $n_\rmn{min}$ and $R_v$ and is not satisfactorily accounted for by either variable alone.} 
\label{fig:Delta_comp}
\end{figure*}

A quantity of significant cosmological interest is the distribution of mass within galaxy voids, which can be determined by measuring the spherically averaged stacked matter density profile about void locations. Previous studies have generally attempted to describe the variation of the stacked profile with the mean size of the voids included in the stack. \citet{Barreira:2015} have proposed a functional form with five free parameters,
\beq
\label{eq:newprofile}
\frac{\rho(r^\prime)}{\overline{\rho}}=1+\delta_c\left(\frac{1-\left(r^\prime/s_1\right)^\alpha}{1+\left(r^\prime/s_2\right)^\beta}\right)\,,
\eeq
where $r^\prime = r/R_v$, to describe the resulting profiles. In Appendix~\ref{appendixA}, we show that this function can provide a satisfactory fit to the stacked profile data for all void sizes $R_v$, and examine similar fitting forms proposed in other studies \citep{Hamaus:2014a,Sutter:2014b,Nadathur:2015a}. Unfortunately the behaviour of the fitted parameters with $R_v$ differs across the different mock galaxy samples, meaning that eq.~\ref{eq:newprofile} is not predictive, i.e., the best-fit parameters must be determined on a case-by-case basis for voids in each sample.

In any case, describing the best-fit profile parameters as a function of $R_v$ alone is not a useful strategy, since it obscures the wide variation in tracer densities within voids of the same size that can be seen in Figure~\ref{fig:bananas}. Indeed \citet{Nadathur:2015b} showed explicitly that the mass density profile of voids is also strongly dependent on the tracer $n_\rmn{min}$ values even for fixed $R_v$. To demonstrate the full variety of watershed void properties, in the next section we consider the dependence on $n_\rmn{min}$ and $R_v$ simultaneously.

\begin{figure*}
\includegraphics[width=145mm]{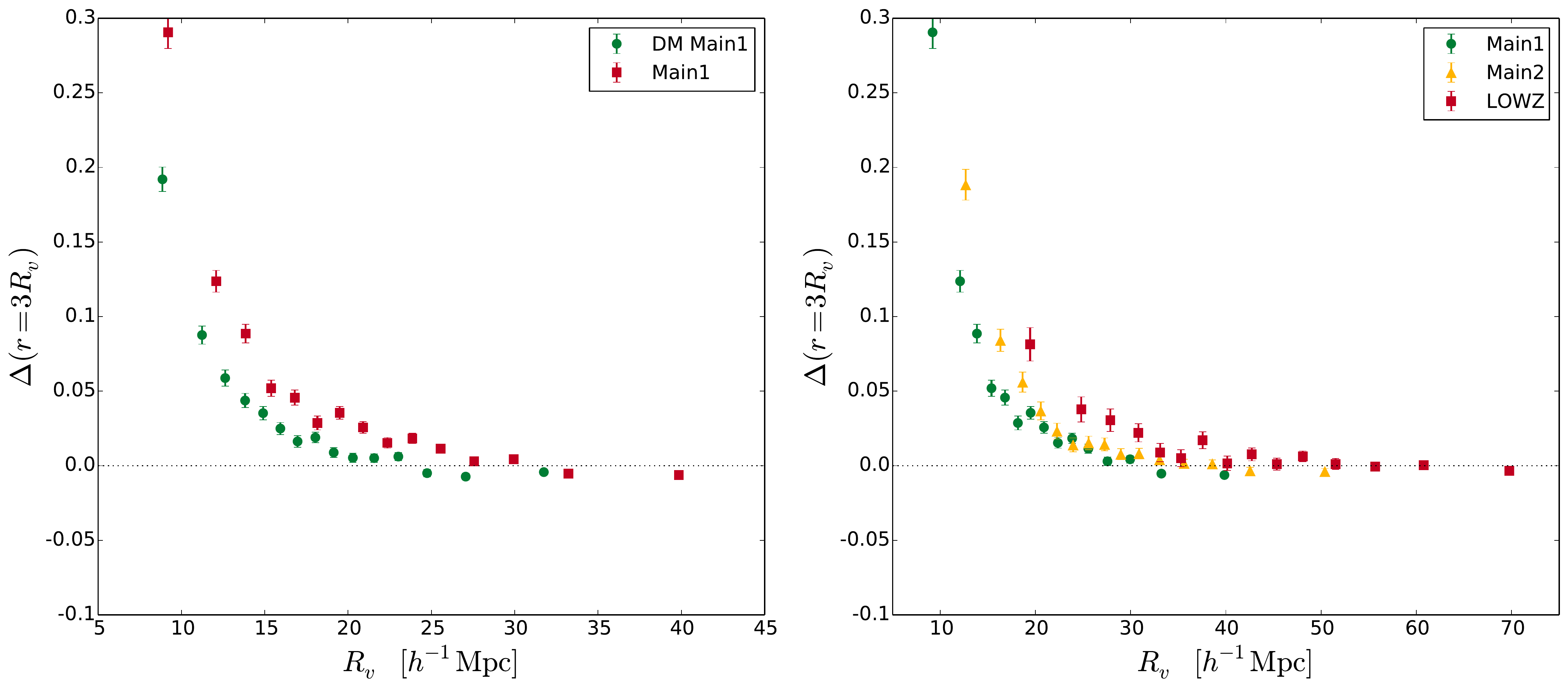}
\caption{The mean enclosed matter density contrast $\Delta(r=3R_v)$ for voids as a function of the void radius $R_v$ alone. \emph{Left}: Comparison between voids in the Main1 HOD sample and those in the unbiased DM Main1 tracers, which have the same mean tracer density $\overline{n}$. Similar results hold for other pairs of samples with the same tracer density. \emph{Right}: Voids in the three HOD mock galaxy samples. In all cases, small voids are on average overcompensated ($\Delta>0$) and large voids on average exactly compensated ($\Delta\simeq0$). The radius scale for the approach to compensation depends on both $\overline{n}$ and the tracer bias. Information on undercompensated voids seen in Figure \ref{fig:Delta_comp} is lost when through averaging over $n_\mathrm{min}$. } 
\label{fig:Rv_Delta}
\end{figure*}

\subsection{Compensation of mass around voids}
\label{subsec:compensation}

Describing the full variation of the density distribution around void centres in terms of a fitting formula such as eq.~\ref{eq:newprofile} and the dependence of its free parameters on void properties is a complicated task. Instead we choose to describe the density environment around the void location by a single number, namely the total enclosed mass density contrast $\Delta$ within a sphere of radius three times the effective void radius $R_v$. This radius is large enough that $\Delta$ undergoes no further zero-crossings at larger $r$. Therefore the sign of $\Delta(r=3R_v)$ distinguishes between two important classes of void environments. Overcompensated voids, with $\Delta>0$, are local density minima embedded within larger scale contracting overdensities, reminiscent of the void-in-cloud scenario described by \citet{Sheth:2003py}. On the other hand, for $\Delta<0$ the interior mass deficit of the void is not compensated by surrounding high density regions. Such voids will correspond to regions of positive gravitational potential, $\Phi>0$. Being able to distinguish between these two cases on the basis of primary void observables is important for studies of the gravitational effects of voids, particularly measurements of lensing and secondary CMB anisotropies due to voids.

Figure \ref{fig:Delta_comp} shows the dependence of $\Delta(r=3R_v)$, measured from the full resolution simulation output, on $n_\rmn{min}$ and $R_v$ for voids in each of the three mock galaxy samples. The same trends are seen across all samples, suggesting the possibility of a universal description of the void environment. However, contrary to previous suggestions \citep{Ceccarelli:2013,Cai:2013ik,Hamaus:2014a} the transition from over-compensation to under-compensation is not a function of the void size alone: both the smallest and largest values of $\Delta$ are found at small void radii. Shallower voids are on average more likely to be over-compensated, so a cut based on the minimum tracer density at roughly $n_\rmn{min}/\overline{n}\lesssim0.3$ \citep[as applied by][]{Hotchkiss:2015a} would better distinguish under-compensated voids, but is still not completely satisfactory. 

\begin{figure*}
\includegraphics[width=150mm]{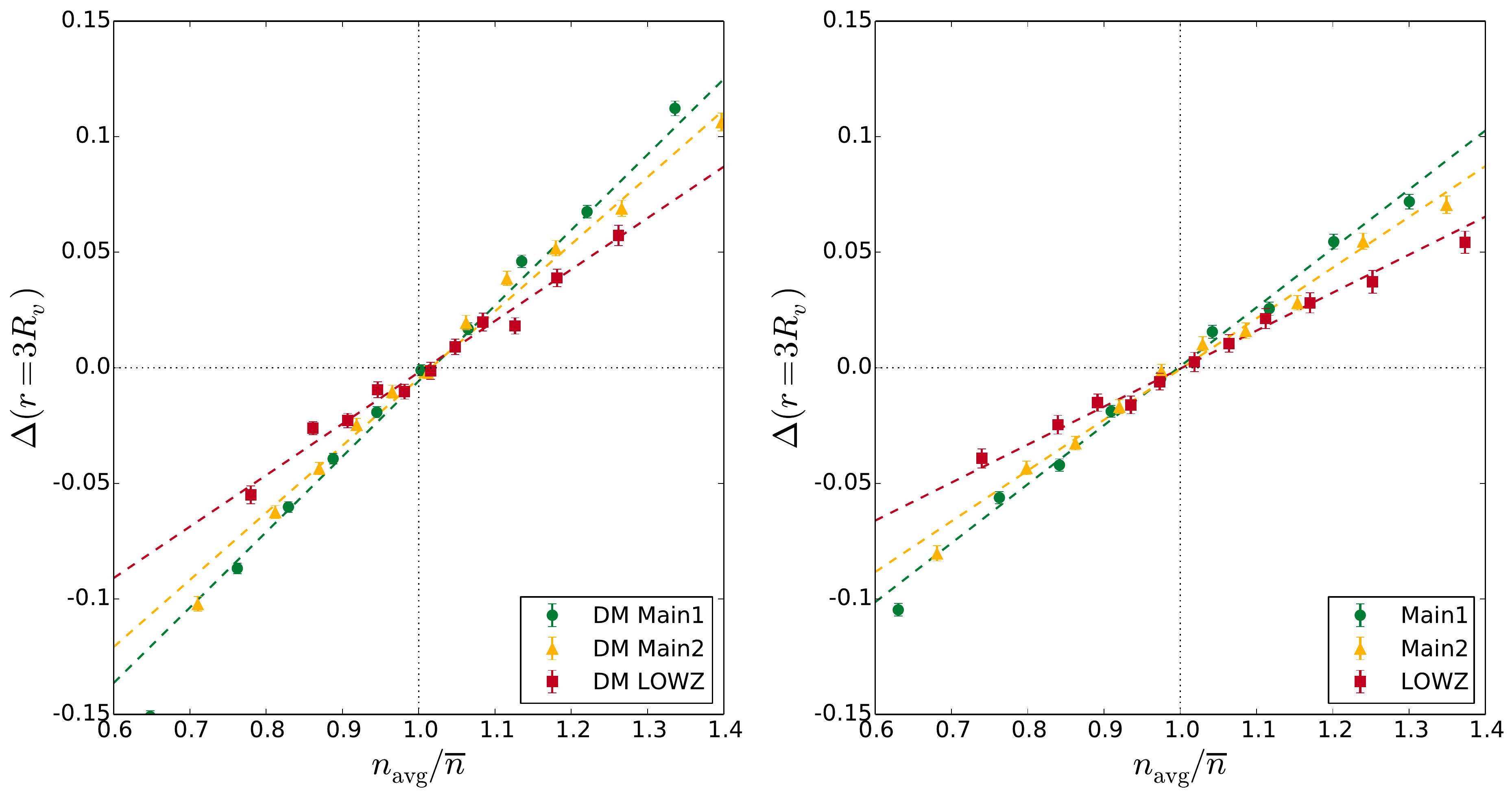}
\caption{Void compensation as measured by the enclosed matter density contrast $\Delta(r=3R_v)$ from eq.~\ref{eq:Delta}, as a function of the overall density of tracers within the void, $n_\rmn{avg}$. The left panel shows the trends for voids in subsampled dark matter, and the right panel for voids in galaxy tracers. The dashed lines show the best fits to the data; in all cases a simple linear relationship provides an excellent description. The value $n_\rmn{avg}=\overline{n}$ provides a universal distinction between under-compensated and over-compensated voids, irrespective of the properties of the tracer population in which voids are identified.} 
\label{fig:navg_Delta}
\end{figure*}

The dependence of void properties on variables other than the void radius is significant for their theoretical modelling. Thus far, such models have followed \citet{Sheth:2003py}, who develop an excursion set model roughly analogous to that for halos. In this scenario, the void radius $R_v$ is analogous to the halo mass and so all void properties, an in particular, the variation of the average density profile, are described purely with reference to $R_v$. Figure~\ref{fig:Delta_comp} shows that this analogy does not work well, because it fails to account for the strong systematic dependence on $n_\mathrm{min}$. In a companion paper \citep{Nadathur:2015b}, we have also shown that watershed voids fail to satisfy several other important assumptions of the \citet{Sheth:2003py} model.

It is however possible to ignore the dependence on $n_\mathrm{min}$ and describe the average behaviour of $\Delta$ as a function of $R_v$ alone, as has been done in previous works. The resulting behaviour is shown in Figure~\ref{fig:Rv_Delta}. For all samples, small voids are on average overcompensated ($\Delta>0$), but approximate compensation is reached at large $R_v$. Note that all information of the undercompensated voids clearly visible in Figure~\ref{fig:Delta_comp} is lost in marginalizing over $n_\mathrm{min}$. The radius scale above which voids are on average compensated is dependent on both the tracer density and bias.

In Figure~\ref{fig:navg_Delta} we show the dependence of $\Delta$ instead on the average tracer number density within a void, $n_\rmn{avg}$, for all six tracer populations used in this study. {Using $n_\mathrm{avg}$ as the independent variable preserves information on both under- and overcompensated voids.} In all cases a simple linear relationship exists between $n_\mathrm{avg}$ and $\Delta$. Straight-line fits to the data show that $n_\rmn{avg}=\overline{n}$ is the universal dividing line between under-compensated and over-compensated voids in every case. We conclude that, in contrast to $R_v$, $n_\rmn{avg}$ is a universal indicator of the compensation of the mass deficit within a void, independent of the bias properties or sampling density of the tracer population in which the voids are identified.

At one level, this result appears intuitively simple: regions containing a higher than average number density of tracers, whether dark matter particles or galaxies, also correspond to a higher than average matter density. However, note that $n_\rmn{avg}$ is defined for voids of arbitrary --- and generally aspherical --- shape determined by the topology of the tracer density field, whereas $\Delta$ refers to the mass density within idealised spheres. There is also a large difference in scales, since $\Delta$ is measured within a volume 27 times larger than that of the void. 

It is also worth noting that the relationship between \emph{local} tracer and matter densities at the void centres is not independent of the properties of the tracer population, as shown in Figure~\ref{fig:biases}. Voids in different samples trace minima of very different depths. Similarly we were also unable to find any comparable relationship between the locally-measured tracer and matter densities at void locations, i.e. the profiles of $\delta(r)$ are not proportional to those of $\delta_n(r)$ for any tracer type. This simple linear relationship exists only for integrated quantities $n_\rmn{avg}$ and $\Delta$.


\section{Conclusions}
\label{section:conclusion}

The existing theoretical model of voids based on the excursion set formalism of \citet{Sheth:2003py} has previously been shown to be inadequate for describing the properties of voids that can be practically identified in simulations by means of the commonly-used watershed algorithm \citep{Nadathur:2015b}. This means that meaningful predictions for void properties and the cosmological constraints that may be obtained from the use of voids can at present only be obtained through a simulation-led approach, but to be observationally relevant such simulations must aim to replicate the types of voids that would be seen in actual galaxy survey data. 

Our aim in this paper has been to provide a comprehensive examination of the twin roles of galaxy bias and sparsity in determining the properties of voids, and in the process to improve existing treatments in the literature.  To do so we made use of a suite of mock galaxy samples in an $N$-body simulation, which were used as tracer populations for void identification. These void catalogues were compared with equivalent ones obtained using unbiased subsets of the dark matter particles in the simulation as tracers. Although our results were obtained specifically using the popular {\sm ZOBOV} algorithm, the general qualitative features of voids we describe will be applicable to any void-finder based on the watershed principle.

Our first practical result has been to show that, contrary to previous claims \citep{Sutter:2014b,Zivick:2014,Sutter:2015a}, tracer bias plays a crucially important role in determining the overall abundance of voids, their size distributions, and matter content. The differences in such void properties introduced by tracer bias are comparable to or larger than those reported from differences in cosmology \citep{Zivick:2014,Yang:2014,Sutter:2015a,Pisani:2015}. This result is not entirely unexpected, since all of these quantities naturally depend strongly on the clustering properties of the tracer population, which are in turn related to the tracer bias. However, it means that the commonly-used strategy of approximating galaxy voids by simply down-sampling the density of dark matter particles in a simulation is incorrect and that predictions for void observables calibrated on such simulation results will unfortunately bear little or no relation to data obtained from real galaxy surveys. A similar conclusion has been independently reached by \citet{Pollina:2015}.

Our results have also clarified the general nature of watershed voids. Such voids correspond to density minima of widely varying depths typically in the linear regime, rather than the highly non-linear evolved objects postulated by some theoretical models. Most previous studies of voids, from both theoretical and simulation perspectives, have focussed on describing void properties as functions of the void size alone. In reality the properties of voids of the same size show a strong systematic dependence on other factors as well. Our work here and in the companion paper \citet{Nadathur:2015b} has aimed to highlight the important role of tracer number densities in particular as predictors of the matter content of voids. 

From this perspective, the most important results in this paper are shown in Figures \ref{fig:Delta_comp} and \ref{fig:navg_Delta}, which show how the large-scale matter fluctuation around void locations can be predicted from easily measurable properties of the tracer galaxies within the void, irrespective of the detailed clustering properties of the galaxy population.  In particular, the average galaxy number density within the void can serve as a universal diagnostic of the mass compensation on much larger scales.

The nature of this mass compensation is interesting in its own right and as a clue towards the formation and evolution history of voids. However, it is primarily important for the use of voids in cosmology because it serves as a signifier of the value and sign of the gravitational potential $\Phi$ at the void location. For instance, for any observational measurement of the gravitational effects of voids, such as through lensing distortions or secondary CMB anisotropies, one would at a minimum like to exclude voids corresponding with $\Phi<0$. In the past, various authors have attempted to use other directly observable properties of watershed voids, such as the ratio of densities in the void wall and at its centre \citep{Granett:2008ju}, the void size \citep{Cai:2013ik,Planck:2013ISW}, or the minimum galaxy number density within the void \citep{Hotchkiss:2015a} as proxies for the value of $\Phi$. Our result indicate instead that $n_\rmn{avg}$ is a better indicator, and that this holds for voids seen in any galaxy sample. A generalization of this measure to a weighted average density for galaxies with a redshift-varying selection function should also be quite straightforward.

We should at this point introduce a note of caution. Strictly speaking, our results in this paper concern the compensation of mass in voids, and not the potential itself. We have used $\Delta$ as a single number with which to characterize the total large-scale mass deficit at void locations, but the gravitational potential depends also on the details of the density profile and the size of the void, as it is related to an integral over $r$ of $\Delta(r)$. Nevertheless, it remains the case that $n_\rmn{avg}$ is an excellent predictor of $\Phi$, as we will show in forthcoming work.

\section*{Acknowledgements}

SH acknowledges support from the Science and Technology Facilities Council [grant number ST/L000652/1] .

The MultiDark Database used in this paper and the web application providing online access to it were constructed as part of the activities of the German Astrophysical Virtual Observatory as result of a collaboration between the Leibniz-Institute for Astrophysics Potsdam (AIP) and the Spanish MultiDark Consolider Project CSD2009-00064. The MultiDark simulations were run on the NASA's Pleiades supercomputer at the NASA Ames Research Center.

\bibliographystyle{mn2e}
\bibliography{refs.bib}

\appendix
\section{Void density profiles and fitting formulae}
\label{appendixA}

\begin{figure*}
\includegraphics[width=175mm]{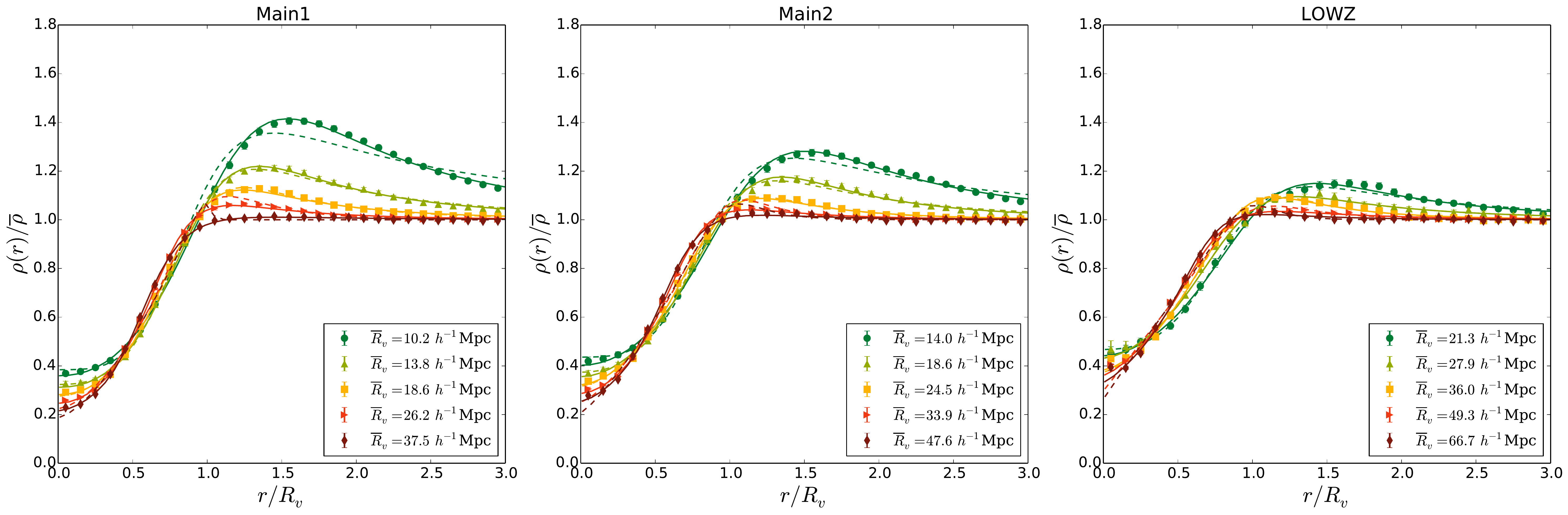}
\caption{Average density profiles for voids of different sizes. Voids in each mock galaxy population are binned on the basis of their radius $R_v$ and stacked so that their centres coincide. The average matter density distribution for the stack is then determined from the dark matter field of the simulation. Solid lines show the best-fit forms of the profile in eq.~\ref{eq:newprofile} in each case, and dashed lines the best-fit forms of eq.~\ref{eq:HSW}. Some radius bins are omitted for clarity.} 
\label{fig:profiles}
\end{figure*}

\subsection{Matter density profiles}
Although stacked void density profiles have been studied by several authors, results in the literature are rather contradictory \citep[e.g][]{Ricciardelli:2014,Hamaus:2014a,Sutter:2014b,Nadathur:2015a,Nadathur:2015b,Barreira:2015}. We start by considering the dependence of the average density profile on the void radius $R_v$. To do this, we create stacks of voids in bins of $R_v$ and centred on the void centre, and determine the average profile $\rho(r)/\overline{\rho}$ for each stack using the gridded density information from the full resolution simulation output. From this data we determine the best-fit parameters $\delta_c$, $s_1$, $s_2$, $\alpha$ and $\beta$ in eq.~\ref{eq:newprofile} for each stack.

Figure~\ref{fig:profiles} shows the profile data for voids of different sizes in each galaxy sample, along with the best-fit forms of eq.~\ref{eq:newprofile} (solid curves) in each case. \citet{Hamaus:2014a} have proposed a `universal' fitting form for the profile, which corresponds to a special case of eq.~\ref{eq:newprofile} with $s_2=1$:
\beq
\label{eq:HSW}
\frac{\rho(r)}{\overline{\rho}}=1+\delta_c\left(\frac{1-\left(r/r_s\right)^\alpha}{1+\left(r/R_v\right)^\beta}\right)\,.
\eeq
The best-fit versions of this constrained profile are shown with dashed lines in Figure~\ref{fig:profiles}. It can be seen that they generally provide somewhat worse fits to the data, both inside the void and in the compensating shell. 

\begin{figure}
\includegraphics[width=80mm]{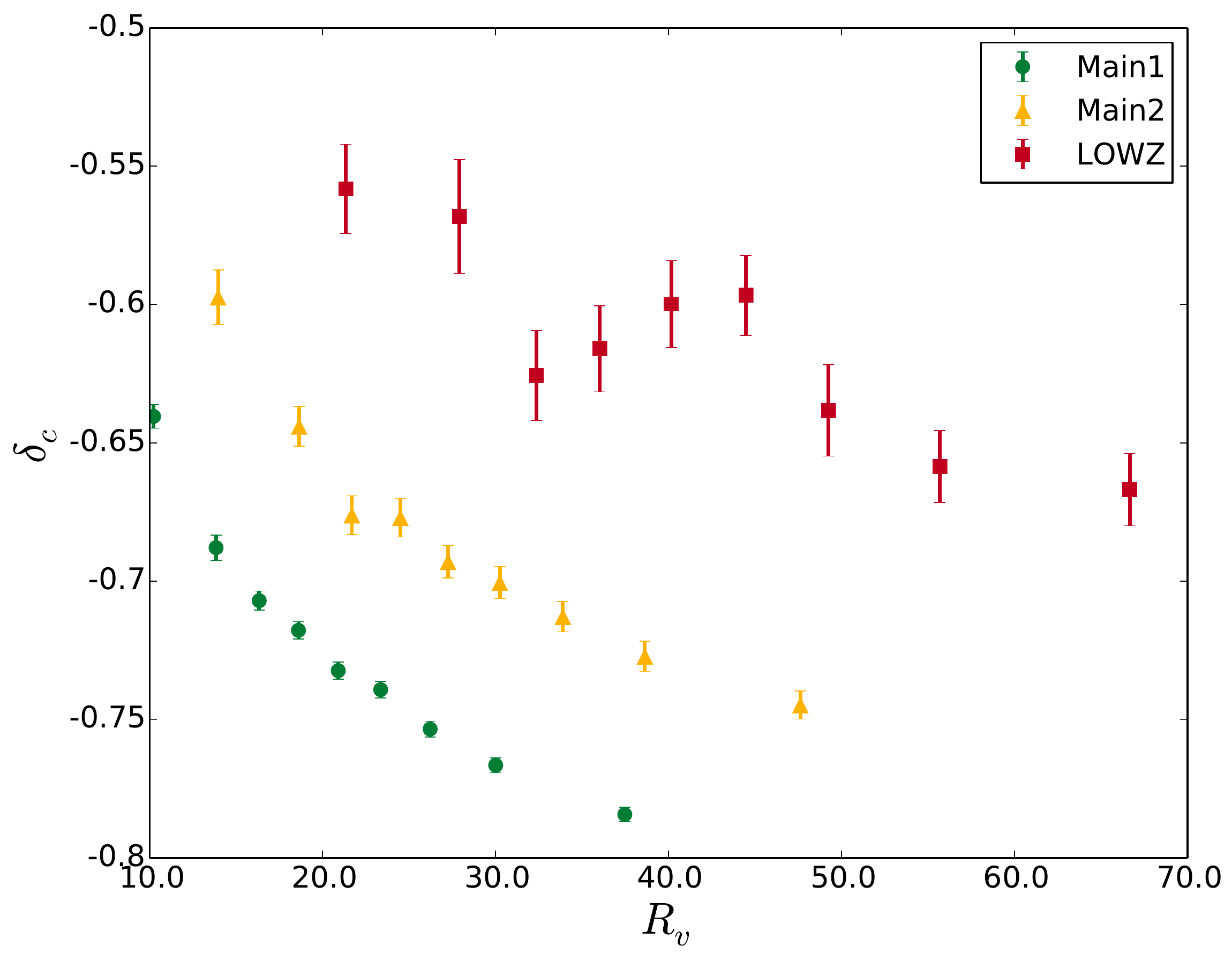}
\caption{The dependence of the best-fit values of the parameter $\delta_c$ in eq.~\ref{eq:newprofile} as a function of the mean void radius of the stacks shown in Figure~\ref{fig:profiles}. In each case $\delta_c$ decreases with $R_v$, and this trend is unchanged if the alternative fitting form in eq.~\ref{eq:HSW} is used, or for stacks centred on the void barycentres.} 
\label{fig:delta_c}
\end{figure}

A key feature of both fits is that, as shown in Figure~\ref{fig:delta_c}, within the sample of voids from each population the fitted parameter $\delta_c$ is a decreasing function of $R_v$. This is consistent with (and can be indirectly deduced from) the generic watershed behaviour shown in Figures~\ref{fig:bananas} and \ref{fig:biases}, and also agrees with the results of \citet{Nadathur:2015b} where we found that  larger voids contain deeper density minima. However, it is opposite to the behaviour claimed in \citet{Hamaus:2014a}, who find $\delta_c$ increases with $R_v$. For the other fitted parameters we find no consistent trends across the different galaxy samples, so we are also unable to confirm the claim of universality of the void density profile made by those authors. In particular, the dependence of $\alpha$ and $\beta$ on scale radius $r_s$ does not follow the pattern described by \citet{Hamaus:2014a} and \citet{Sutter:2014b} (see \citealt{Nadathur:2015b} for more details).

It should be noted that \citet{Hamaus:2014a} proposed eq.~\ref{eq:HSW} as a description of the average density profile for stacks centred about void barycentres, whereas in this paper we define the void centre differently to better identify the true location of the underdensity. In \citet{Nadathur:2015b} we showed that for some high-density tracer populations and for some choices of merging criteria, the average density at the barycentre can show a non-monotonic behaviour with increasing void size (see Figure 3 in particular). This occurs due to the displacement of the barycentre location from the true underdensity in voids composed of very large numbers of merged sub-voids. Even in these situations, however, $\delta_c$ does not increase monotonically with $R_v$ as claimed by \citet{Hamaus:2014a}. This unusual behaviour also does not apply for any of the galaxy samples used in this work, which all show monotonically \emph{decreasing} $\delta_c(R_v)$ irrespective of the choice of void centre.

However, it is worth stressing again that analyzing the variation of the mean density profile as a function of $R_v$ obscures the important sytematic variation across the void population with the values of $n_\rmn{min}$ shown in Figure~\ref{fig:Delta_comp} and discussed in \citet{Nadathur:2015b}. The stacked profiles in $R_v$ bins in this Appendix are therefore included only to enable comparisons with previous results in the literature. For voids of the same size, the variation of the stacked density profile in stacks of the same $n_\rmn{min}$ is large, as expected from Figure~\ref{fig:Delta_comp}. If voids are stacked according to their $n_\rmn{min}$ or $n_\rmn{avg}$ values rather than $R_v$,  the form of eq.~\ref{eq:HSW} is completely unable to fit the resulting profiles. However, the form of eq.~\ref{eq:newprofile} continues to provide a good description in all cases.

\subsection{Galaxy number density profiles}

\begin{figure}
\includegraphics[width=80mm]{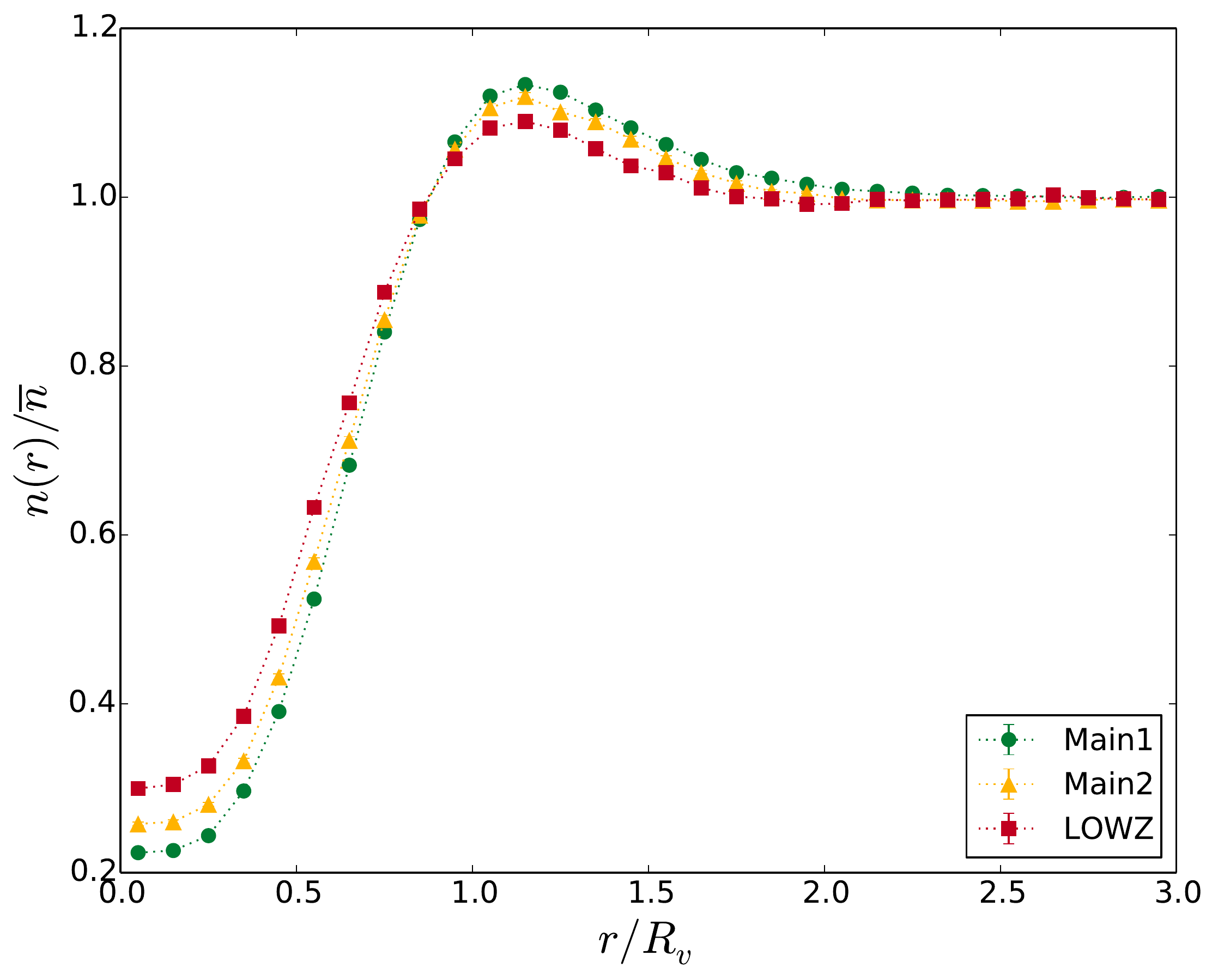}
\caption{Galaxy number density profiles for voids in the three mock galaxy samples. To match the methodology used by \citet{Nadathur:2015a} a selection cut  $n_\rmn{min}/\overline{n}<0.3$ is applied and stacks are centred on void barycentres. The resulting average profiles in the different galaxy populations are similar but statistically distinguishable given the large sample size. The dotted lines are linear interpolations and do not represent fits to the data.} 
\label{fig:non-universality}
\end{figure}

In a previous work \citep{Nadathur:2015a}, we found that the stacked galaxy number density profile for a selected subset of voids in several different galaxy populations drawn from data from SDSS surveys at different redshifts were very similar, and argued that this might be a sign of universality. However, the size of the data samples used in this work were small, and the errors in determination of the profiles were correspondingly large. The number of voids in our simulation samples exceeds that used in the previous work by a factor of $\sim10$: this is because the simulation volume is much larger than the survey volume for some of the observational samples, and boundary effects due to survey masks are not present. We are therefore able to re-examine the universality of galaxy density profiles with greater statistical precision.

\citet{Nadathur:2015a} analysed only those voids with $n_\rmn{min}/\overline{n}<0.3$. In the absence of information of the true matter content of galaxy voids, this very conservative cut was intended to remove Poisson contamination of the void sample, based on calibration of the probability $P(n_\rmn{min}|\rmn{Poisson})$. \emph{Ad hoc} cuts to the void population on the basis of void radius $R_v$ have been advocated elsewhere in order to achieve the same purpose. Our analysis of the matter density content of voids in simulation, both in this paper and in \citet{Nadathur:2015b}, indicates that on average all voids  correspond to true matter underdensities, so such cuts are not necessary and lead to sub-optimal use of available data. However, for the purposes of direct comparison with the results of \citet{Nadathur:2015a}, we apply the same methodology here. We then stack the surviving voids in each of the Main1, Main2 and LOWZ mock galaxy samples separately, without binning by radius, to determine the average galaxy number density profiles. For consistency, we also use the 'VTFE' estimator for measuring galaxy density profiles discussed in that work and centre stacks on the void barycentres, though the results are not strongly dependent on this choice. The resulting profiles are shown in Figure~\ref{fig:non-universality}. The results in the different galaxy populations are similar to each other, but distinguishable with high statistical significance due to the small error bars. However they are also consistent with the profile seen in \citet{Nadathur:2015a} within the much larger statistical errors of that study.

\label{lastpage}
\end{document}